# Relativistic Positron-Electron-Ion Shear Flows and Application to Gamma-Ray Bursts

Edison Liang[1], Wen Fu[1], Markus Boettcher[2,3], Ian Smith[1], Parisa Roustazadeh[1,3]

[1] Rice University, Houston, TX 77005

[2] Centre for Space Research, North-West University, Potchefstroom, 2520, South Africa

[3] Ohio University, Athens, OH 45701

ABSTRACT

We present Particle-in-Cell simulation results of relativistic shear flows for hybrid positron-electron-ion plasmas and compare to those for pure e+e- and pure e-ion plasmas. Among the three types of relativistic shear flows, we find that only hybrid shear flow is able to energize the electrons to form a high-energy spectral peak plus a hard power-law tail. Such electron spectra are needed to model the observational properties of gamma-ray bursts.

Subject Keywords: Shear Flow; Gamma-Ray - Bursts

## I. INTRODUCTION

Recent large-scale Particle-in-Cell (PIC, Birdsall & Langdon 1991) simulations of relativistic shear boundary layers (SBLs) demonstrated efficient generation of ordered magnetic fields and particle energization from initially unmagnetized plasmas (Alves et al 2012, Liang et al 2013, Grismayer et al 2013). In addition to blazar jets (Boettcher 2007, Ghisellini et al 2005), relativistic SBLs are likely to occur in gamma-ray burst (GRB) jets, which involve ultra-relativistic outflows from stellar-mass collapsed objects interacting with stationary circumstellar and interstellar media (Piran 2000, 2004, Meszaros 2002). Previous PIC simulations of SBLs focussed on pure leptonic (e+e- pairs, Liang et al 2013) or pure e-ion flows with low bulk Lorentz factors (Alves et al 2012, Grismayer et al 2013). Here we generalize the previous works to include hybrid e+e-ion plasmas at high Lorentz factors ($p_o$ =15, where $p_0$ is the initial bulk flow Lorentz factor in the center of momentum (CM) frame). We find that only hybrid SBL can accelerate electrons to form a broad high-energy spectral peak followed by a hard power-law tail, resembling the typical observed GRB spectrum (Preece et al 2000, Meegan et al 2009). In this paper we summarize the PIC simulation results and discuss their potential applications to the optically thin synchrotron (OTS) model of GRB emissions (Piran 2005, Meszaros 2002). Our simulations are performed in the CM frame in contrast to some MHD simulations which use the laboratory frame of the central engine (Mizuno et al 2007).

We first summarize the results of pure e+e- and pure e-ion relativistic SBLs. In the pure e+e- case (Liang et al 2013), transverse electromagnetic (EM) fields are first generated by oblique Weibel-like (Weibel 1959, Yoon 2007, Yang et al 1993, 1994) and 2-stream (Boyd & Sanderson 2003, Lapenta et al 2007) instabilities driven by interface-crossing counter streams. The small-scale fields then self-organize into larger and larger ordered quasi-periodic magnetic flux ropes and electric field channels, with alternating polarity. Later they coalesce into large dipolar EM "vortices" extending over hundreds of skin depths (=$c/\omega_e$, $\omega_e$=electron plasma frequency). The ultimate sizes of these dipolar EM vortices are likely limited by the kinetic energy inflow rate versus turbulence dissipation rate. Particles are energized by cross-field drift motions (Boyd & Sanderson 2003), leading to a nonthermal soft quasi-power-law which turns over at ~$p_o/2$. A density trough is formed at the SBL due to plasma expulsion by the extra magnetic pressure. The asymptotic field energy is ~ 8% of the total energy (Liang et al 2013).

For pure e-ion ($m_i/m_e$=1836) shear flow, the properties of the SBL are drastically different. In the $p_o \leq 3$ cases studied by Alves et al (2012) and Grismayer et al (2013), the initial instability is a kinetic version of the Kelvin-Helmohltz instability (KHI, Chandrasekhar 1981) with free-streaming ions and fluid-like electrons. We have extended the pure e-ion runs to $p_o=15$ using much larger boxes. As we see below in Sec.II, current sheets are formed on opposite sides of the interface, sustaining a steady slab of monopolar magnetic field. The strong magnetic pressure expels the plasma, forming a density trough at the interface, but the ions are expelled more than the electrons due to larger ion gyroradii. This leads to charge separation and formation of a triple layer, with electric fields pointing from both current sheets towards the center of the SBL. Electrons are efficiently accelerated by cross-field drift motions, eventually reaching the ion energy. At late times the electron distribution forms a narrow peak at the decelerated ion energy $\sim p_o m_i c^2/2$ with sharp cutoff: there is *no evidence of any power-law tail beyond the peak*. The asymptotic field energy is $\sim 12\%$ of the total energy (see Sec.II).

## II. PIC SIMULATION RESULTS

We used two PIC codes in this work: the 2.5D (2D space, 3-momenta) Zohar from LLNL (Birdsall & Langdon 1991, Langdon & Lasinski 1976) and the 3D EPOCH from the University of Worwick UK, the academic open source version of the code PSC (Ruhl 2006). We compared results from both codes for many different runs and with Alves et al (2012) and validated their agreement. We also checked that the relativistic Cerenkov and other numerical instabilities (Birdsall & Langdon 1991) are well suppressed in both codes: total energy is conserved to < 0.1% in Zohar runs and few % in EPOCH runs. Fig.1(a) shows the problem setup in the CM frame. 2D runs in the x-y plane are referred to as P-mode and 2D runs in the y-z plane are referred to as T-mode (Liang et al 2013). We have studied both modes, but found that the *T-mode instabilities grow very slowly compared to the P-mode for $p_o=15$ when ion energy dominate*s. 3D runs also show that gross SBL properties and the asymptotic electron spectra are similar to 2D P-mode (see below). While we will show some 3D results for comparison (Fig.1(b)

Inset & Fig.4), *this paper will focus on the 2D P-mode* results since we can use much larger boxes in 2D. We use doubly periodic x-y boxes ranging from 1024x2048 (Zohar) to 1024x12288 (EPOCH) with 20-40 particles per cell. The larger EPOCH box allows us to run much longer without interaction between the two shear layers. The initial shear Lorentz factor $p_o$=15 in the CM frame, much higher than the cases studied by Alves et al (2012). The plasma is *initially unmagnetized*. The initial temperature T= 2.5keV for electrons, positrons and ions ($m_+/m_e$=1, $m_i/m_e$=1836, ions carry almost all initial energy). The initial electron density n is normalized to 1 so that cell size = plasma skin depth. We set time step=$0.1/\omega_e$ so that Boris rotation steps remain stable (Birdsall & Langdon 1991) even in the strongest field regions. Spatial distances are measured in $c/\omega_e$ and time is measured in $1/\omega_e$. All physical quantities refer to the CM frame.

We compare the SBLs of three different plasmas: (A) pure e-ion, (B) 90% e-ion +10% e+e- (hybrid case), (C) pure e+e- (from Liang et al 2013). Figs.1(b)(c)(d) compare the time evolution of the different energy components for the three different plasmas. At late times, the field energy saturates at ~12% of total energy for case (A) and ~ 8% for cases (B), and (C) if we sum the P-mode and T-mode (Liang et al 2013). These saturation levels appear insensitive to the box size and are much higher than MHD results (Zhang et al 2009). Figs.1(b)(c) show that the lepton energy eventually reaches equipartition with the ion energy, but case (B) takes longer than case (A). However the e-ion equipartition time scales ~linearly with y-dimension, suggesting that effective e-ion coupling via plasma instabilities begins at the SBL but eventually reaches the whole box. Thus *relativistic SBLs efficiently convert bulk flow energy into EM fields, which in turn energize the leptons and radiate efficiently*. In Fig.1(b) Inset shows the energy histories of our largest 3D (128x2048x64) e-ion run. Compared to 2D P-mode we note that the 3D field energy is higher (> sum of 2D P and T modes), lepton energy is lower and the leptons take longer to reach equipartition with the ions. Some of these differences are caused by the nonlinear excitation of oblique modes, but some may also be due to the smaller 3D box and numerical effects. The true 3D physical effects can only be settled using much larger runs (e.g.1024x2048x1024) with higher fidelity, which is beyond the scope of this paper. Fig.2 compares the $\mathbf{B}_z$ and $\mathbf{E}_y$ profiles (the dominant components) for all three plasmas and Fig.3 compares the $\mathbf{J}_x$, net charge $n_+-n_-$ and vertical density profile n(y) (averaged over x) for cases (A)(B) at $t\omega_e=3000$. As in Alves et al (2012), we find that pure e-ion shear flows create a monopolar slab of uniform $\mathbf{B}_z$ along the shear interface, sustained by $\mathbf{J}_x$ currents due to different electron and ion drift velocities and densities (Fig.3cf), in contrast to the dipolar EM "flywheels" of the pure e+e- shear flows (Fig.2cf, Liang et al 2013). Plasma is expelled from the shear interface by the extra magnetic pressure, creating a density trough (Fig.3cf). However, in cases (A)(B), the leptons, with smaller gyroradii, are less displaced than the ions, leading to charge separation and formation of a triple layer (Fig.3cf), with the electric fields pointing from both ion-dominated current sheets towards the electron-dominated center (Fig.2de). The cross $\mathbf{E}, \mathbf{B}$ fields efficiently accelerate the leptons. The hybrid e+e-ion SBL is different from both the pure e-ion and pure e+e- SBL: outside the hybrid triple layer which is thinner than the pure e-ion triple layer, wavy EM fields (Fig.2be) are created which provide additional channels of particle acceleration.

In Fig.4(a) we compare the electron spectra f(γ) (γ=Lorentz factor) of the three different cases at $t\omega_e=3000$. The pure e+e- spectrum peaks at $\gamma\sim p_o/2$ with a soft high-energy tail produced by stochastic acceleration of vortical EM turbulence (Liang et al 2013). For pure e-ion case (A), the electron spectrum consists of a broad low-energy component plus a narrow high-energy peak at $\gamma\sim p_o m_i/2m_e$ (~decelerated ion energy at late times) with steep cutoff. The electrons in the high energy peak are mainly energized by the triple-layer: the steep cutoff is due to the uniformity of the potential gap across the triple layer (Fig.2d), and the absence of nonthermal tail above the peak is due to the lack of additional stochastic acceleration by waves. In contrast, for the hybrid e+e-ion case (B), the broad low-energy component, which is mainly energized by wave turbulence (see Fig.5), merges with the (weaker) high energy peak energized by the triple layer to form a quasi-power-law. Fig.4(b) shows the detailed electron spectral evolution for the hybrid case. By $t\omega_e$=45000, not only has the electron peak migrated to the decelerated ion energy $\sim p_o f m_i/2m_e$ (f=ion fraction), a well-defined *power-law tail also formed above the peak with slope ~-3*. For optically thin synchrotron (OTS) and Compton radiation, such electron slope corresponds to a photon index ~ -2 (Rybicki and Lightman 1979). In Fig.4(c) we compare the pure e-ion electron spectra in 3D(128x2048x64) vs. 2D: the 3D peak is broader, but its location and the steep cutoff are identical to the 2D spectrum. The slight broadening is expected since in 3D the triple-layer potential gap is less uniform than in 2D due to oblique mode perturbations. In Fig. 4(d) we show the electron phase plot of the hybrid case at $t\omega_e$=10000: the momentum distribution is highly anisotropic so that synchrotron radiation would be strongly beamed along **x** even in the CM frame. Fig.5 shows the time-lapse $B_z$ evolution of our largest 2D hybrid run. The slab fields at the SBL stop expanding after $t\omega_e$~20000, whereas the wavy fields outside the slab continue to expand over time and coalesce into large alternating dipolar flux tubes.

## III. APPLICATION TO GRB

GRB prompt emission spectra exhibit a broad spectral peak at $E_{pk}$ ~ 0.2-0.5 MeV (Preece et al 2000) plus a power-law tail with photon index β ~ -2 to -2.5 (Band 1993), extending to >10 GeV in many Fermi bursts (Abdo et al 2009). In some Fermi LAT bursts there is also evidence of a second hard component (Ackermann et al 2011, Kocevski et al 2012). Historically, most GRB models invoke OTS emission by power-law electrons (Piran 2005, Meszaros et al 2002) and interpret $E_{pk}$ as the synchrotron critical frequency (Rybicki & Lightman 1979) of the electron peak energy. Recently there was a revival of the Thomson-thick photospheric model (Giannios & Spruit 2008, Liang 1997, Liang et al 1997). Both models have their limitations and may apply to different subclasses of GRB spectra: the former may be more relevant to spectra with soft slope below $E_{pk}$ (Band (1993) $\alpha$ index < -1, cf. Preece et al 2000), while the latter more relevant to spectra with hard slope below $E_{pk}$ (Band $\alpha$ > +1, Preece et al 2000, Crider et al 1997). Here we focus on the OTS interpretation of $E_{pk}$. The Thomson-thick photospheric model will be studied in a later paper. For OTS models, the observed properties of GRBs constrain the emission region parameters (density, magnetic field, size; Piran 2005, Liang and Noguchi 2009). In addition to the internal shock scenario (Piran 2005, Meszaros 2002), SBL is also a strong candidate since SBL naturally arises when a relativistic jet passes through stationary medium. If we apply the pure e+e- SBL spectrum (Fig. 4a) to model the OTS $E_{pk}$ value, we find that it predicts emission parameters orders of magnitudes different from the observed GRB values (too low $E_{pk}$, see below), which rules out the pure e+e- SBL model. If we use the pure e-ion SBL electron spectrum to model $E_{pk}$, it predicts emission parameters consistent with observed GRB values, but it cannot produce any hard power-law tail above $E_{pk}$ (Fig. 4c), which is observed in all GRBs. This dilemma drove us to explore the *hybrid e+e-ion SBL model, which is able to produce both high $E_{pk}$ and a hard power-law tail above the peak (Fig. 4b)*.

The hybrid SBL results show that the electron peak energy in the CM frame lies at $\varepsilon_{pk}$~ $p_o m_i c^2 f/2$, where f = ion fraction, $2p_o^2 = \Gamma$ (=bulk Lorentz factor in the host galaxy frame), and the field energy is ~8% of total energy (Fig.1c). Modeling the detailed GRB spectra from PIC simulations involves synthesizing radiation from many particle tracks (Martins et al 2009, Sironi

& Spitkovsky 2009, Frederiksen et al 2010, Nishikawa et al 2011, 2012), which is beyond the scope of this paper. However, since **B** is not too turbulent (Fig.2) and electrons are mainly accelerated orthogonal to field lines, we are justified to use the classical synchrotron formula for a *rough order-of-magnitude estimate*. The critical synchrotron frequency (Rybicki & Lightman 1979) observed on earth, after boosting from the CM frame to the GRB frame with the Lorentz factor $p_o$ and redshifted by z of the host galaxy, is given by $E_{pk} \sim p_o(\varepsilon_{pk}/m_ec^2)^2 h\nu_e/(1+z)$, where $\nu_e = eB/2\pi mc$ is the electron gyrofrequency in the CM frame. For example, with $p_o = 15$ ($\Gamma = 450$), f = 0.9, z = 1 and $E_{pk} = 250$ keV (Preece et al 2000), the above formula predicts B (magnetic field in the CM frame) $\sim 2\times10^4$ G. Using $B^2/4\pi p_o nm_ic^2 \sim 8\%$, we obtain ion ($\sim$electron) density n (in the CM frame) $\sim 10^{10}/cm^3$. These predicted (n, B) values are consistent with those deduced empirically from GRB observations (Piran 2000, 2005, Liang & Noguchi 2009). In contrast, if we had used the pure e+e- SBL electron peak energy $\varepsilon_{pk} \sim p_o m_e c^2/2$, the predicted (n, B) values would be *orders of magnitude off* the commonly accepted GRB values. Hence only ion-dominated SBLs can energize electrons to a sufficiently high electron peak energy required by the OTS model. But *we also need a finite e+e- component to generate additional EM wave turbulence outside the SBL*, to stochastically accelerate electrons to form the hard power-law above the peak. This result demonstrates the power of kinetic simulations for relativistic plasmas. Another question is the effect of radiation damping on particle acceleration in the current model. For $n \sim 10^{10}/cm$, $\omega_e \sim 10^{10}$ rad/sec. Hence $t\omega_e = 45000$ (Fig.4c) corresponds to t $\sim 4\times10^{-6}$ sec. For $B \sim 10^4$ G and $\gamma_{max} \sim 5\times10^4$ (Fig.4c), the synchrotron cooling time for electrons is $\sim 2\times10^{-4}$ sec, or $\sim 50$ times longer than the acceleration time. Thus we are justified in ignoring radiation damping for the PIC simulations presented here. For much higher $\gamma$ and/or B, we may need to include radiation damping effects (Jaroschek & Hoshino 2009, Noguchi et al 2005).

Realistic GRB scenarios likely involve hybrid e+e-ion jets since the central engine is so hot that pairs should be copiously created (Piran 2005), and the jet likely entrains large amount of baryonic matter while emerging from the collapsar. Observationally, a small pair fraction may be difficult to detect directly. But if the pair fraction is large, Faraday rotation effects on polarized radiation may be detectable since positrons and electrons induce opposite rotations. We have only scratched the tip of the hybrid SBL parameter space. Besides varying f and $p_o$, future simulations will explore the effects of density and composition jumps across the shear interface and embedded primordial magnetic fields in the jet.

This work was supported by NASA Fermi Cycle 4 & 5 grants NNX12AE31G and NNX12AO81G. We thank Prof. Arber of the University of Worwick, UK, for providing the EPOCH code. EPOCH runs on Rice clusters were supported by the Data Analysis And Visualization Cyberinfrastructure funded by NSF grant OCI-0959097. ZOHAR runs were performed at Lawrence Livermore National Laboratory.

Figure Captions

Fig.1 (a) Setup of the shear flow simulations. We focuse on the 2D P-mode in the x-y plane. In all figures spatial scales are in units of $c/\omega_e$. (b) 100% e-ion run energy histories; (c): 90% e-ion, 10% e+e- hybrid run energy histories; (d) 100% e+e- run energy histories. In (b)(c), we plot the time histories of the field energy ($E_{em}$), lepton energy ($E_e$) and ion energy ($E_i$) for the P-mode only (T-mode effects are too small to be visible here). In (d), we plot the field energy ($E_{em}$) and lepton energy ($E_e$) for the P-mode and T-mode separately (from Liang et al 2013). Fig.1(b) Inset : energy histories for 3D (128x2048x64) e-ion run.

Fig.2 *Left column*: $B_z$ (in and out of the plane) contour plots with $p_o$=15 at $t\omega_e$=3000: (a) 100% e-ion; (b) 90% e-ion, 10% e+e-; (c) 100% e+e-. The transition from a monopolar slab pattern to dipolar vortical "flywheel" pattern is evident. *Right column*: $E_y$ contour plots for the same three cases showing the transition from a flat triple layer in pure e-ion case (d) to triple layer plus EM waves in hybrid case (e) to oblique electric channels in pure e+e- case (f) (Liang et al 2013). Color bar refers to both left and right figures. Units are arbitrary.

Fig.3 *Left column*: panels (a,b,c) are for pure e-ion SBL. *Right column*: panels (d,e,f) are for hybrid e+e-ion SBL. All figures refer to $t\omega_e$=3000. Figures (a,d) show the current distribution $J_x$. Stronger outer current sheets are dominated by ions, weaker inner current sheets are dominated by electrons. Panels (b,e) show the net charge distribution $n_+$-$n_-$. The hybrid triple layer in (e) is thinner than the pure e-ion triple layer in (b). Color bar refers to figure to the left. Units are arbitrary. Panels (c,f) are the x-averaged vertical density profiles n(y). For panel (c), the labels are: 1. electrons, 2. ions, 3. net charge density. Initial densities are normalized to 0.001. For panel (f), the labels are: 1. electrons, 2. ions, 3. positrons, 4. net charge density. Initial densities

are normalized to 0.001 for electrons, 0.0001 for positrons and 0.009 for ions. In both cases the ions are completely evacuated at the SBL, but the leptons are not, creating net charge separation and formation of the triple layer. The positrons are largely expelled from the ion-dominated current sheets (spikes).

Fig.4 (a) Comparison of the electron spectrum f(γ) for the three different plasmas at $t\omega_e$=3000. Spectrum for pure e+e- plasma (dashed) has a broad peak at ~$p_o$/2 plus a soft nonthermal tail (Liang et al 2013). Electron spectrum for pure e-ion plasma (solid) shows a narrow electron peak at ~$p_o m_i/2m_e$ with steep cutoff. Both spectra show little further evolution for $t\omega_e$ >> 3000. In contrast, electron spectrum for the hybrid case (dot-dashed) exhibits a broad peak with nonthermal high energy tail, both of which migrate towards higher energy as time progresses. (b) Electron spectral evolution for the hybrid case showing a broad peak at ~$p_o f m_i/2m_e$ plus hard power law of slope ~ -3 above the peak at $t\omega_e$=45000. (c) Comparison of 3D electron spectrum with 2D at $t\omega_e$=6000 for e-ion run (vertical scales not normalized). (d) $p_x$ vs. $p_y$ phase plot for the hybrid case at $t\omega_e$=10000 for initially left-moving electrons.

Fig. 5 Time-lapse snapshots of **$B_z$** contour plots for the (1024x12288) hybrid e+e-ion run (color bar has been renormalized between different panels). The larger y-axis allows the problem to run much longer without interaction between the two SBLs.

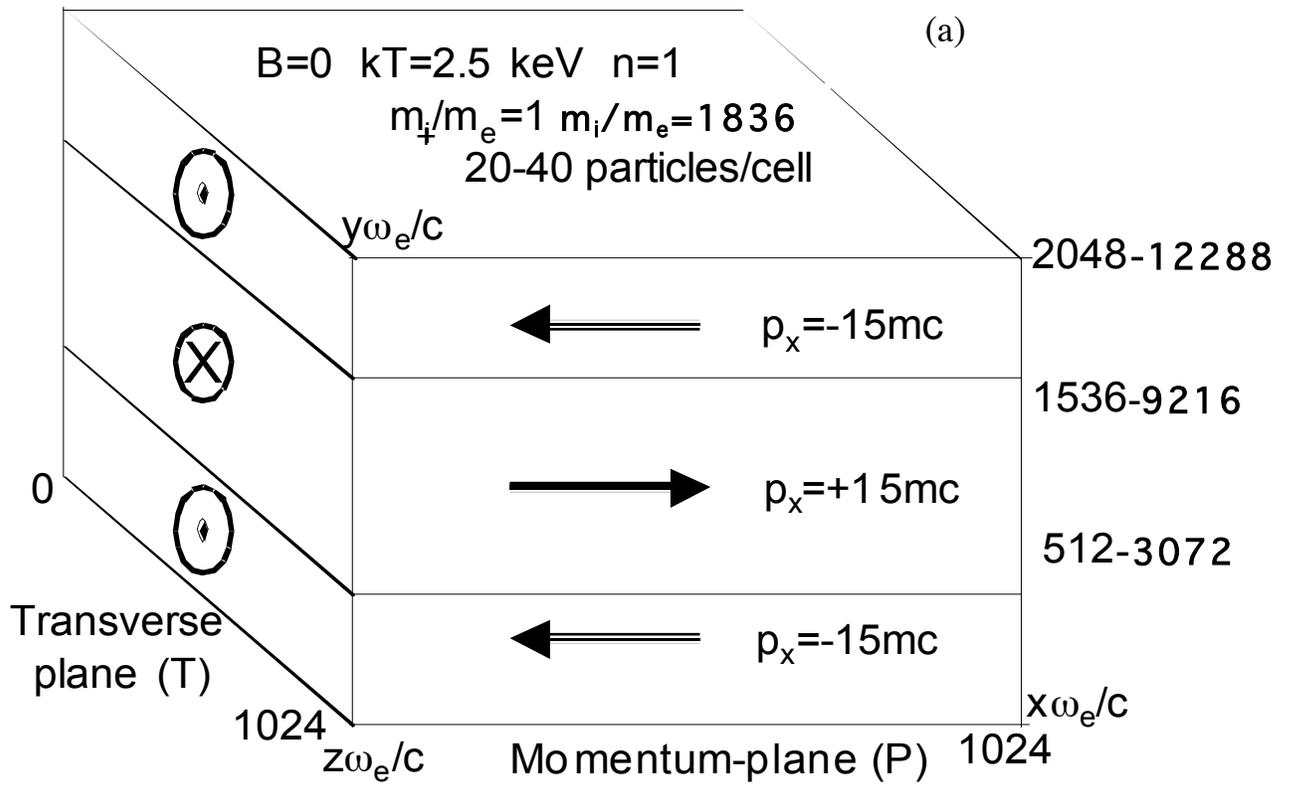
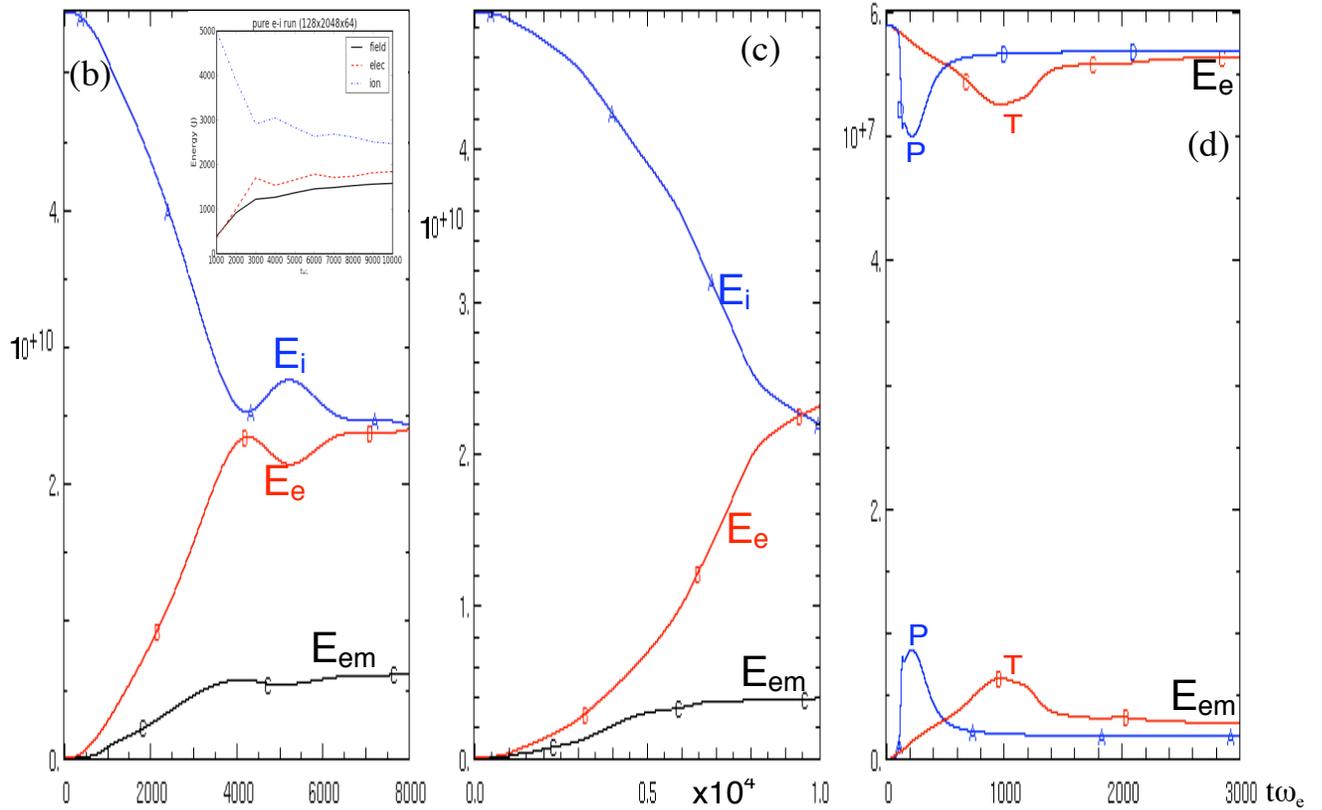

Fig.1

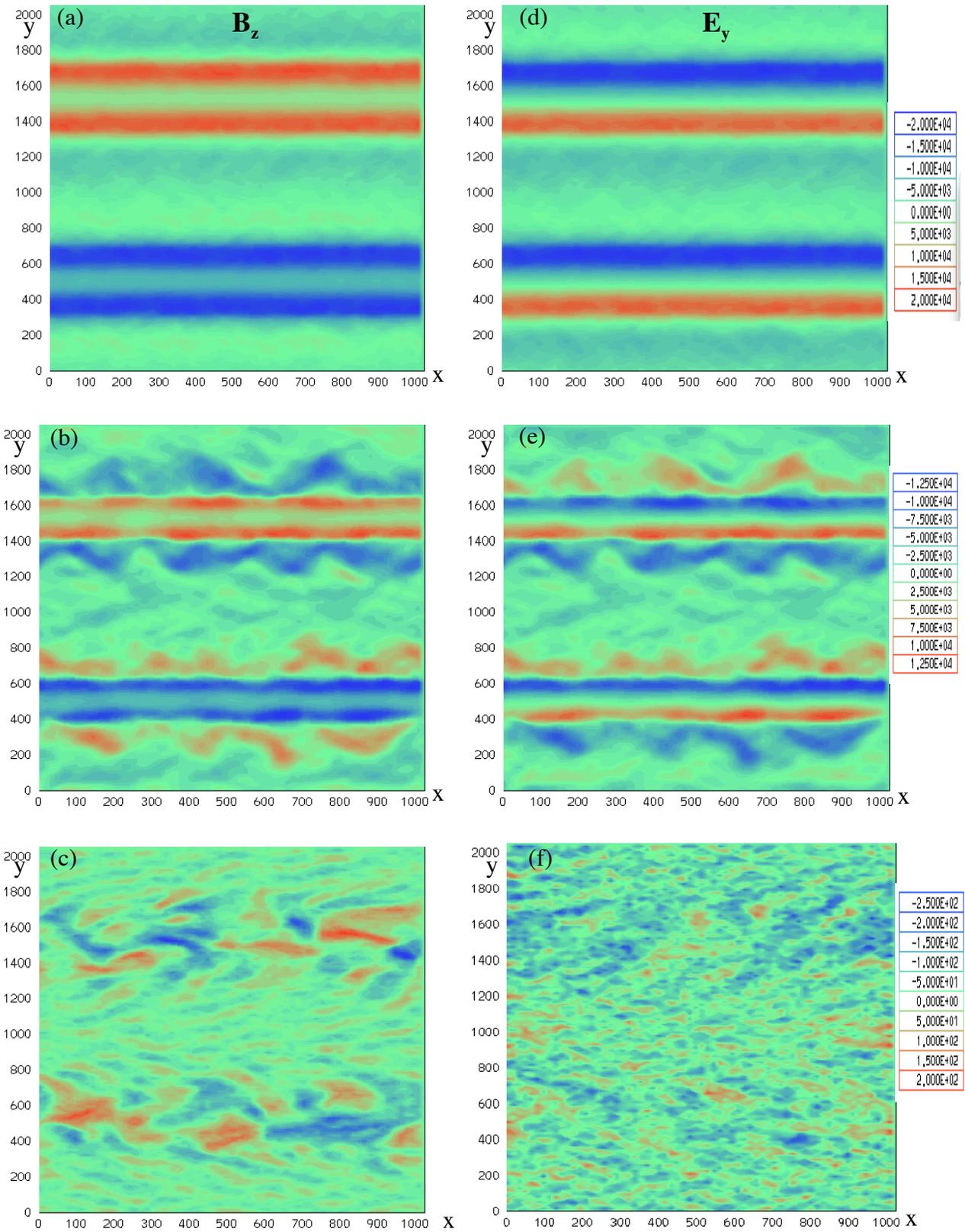

Fig.2

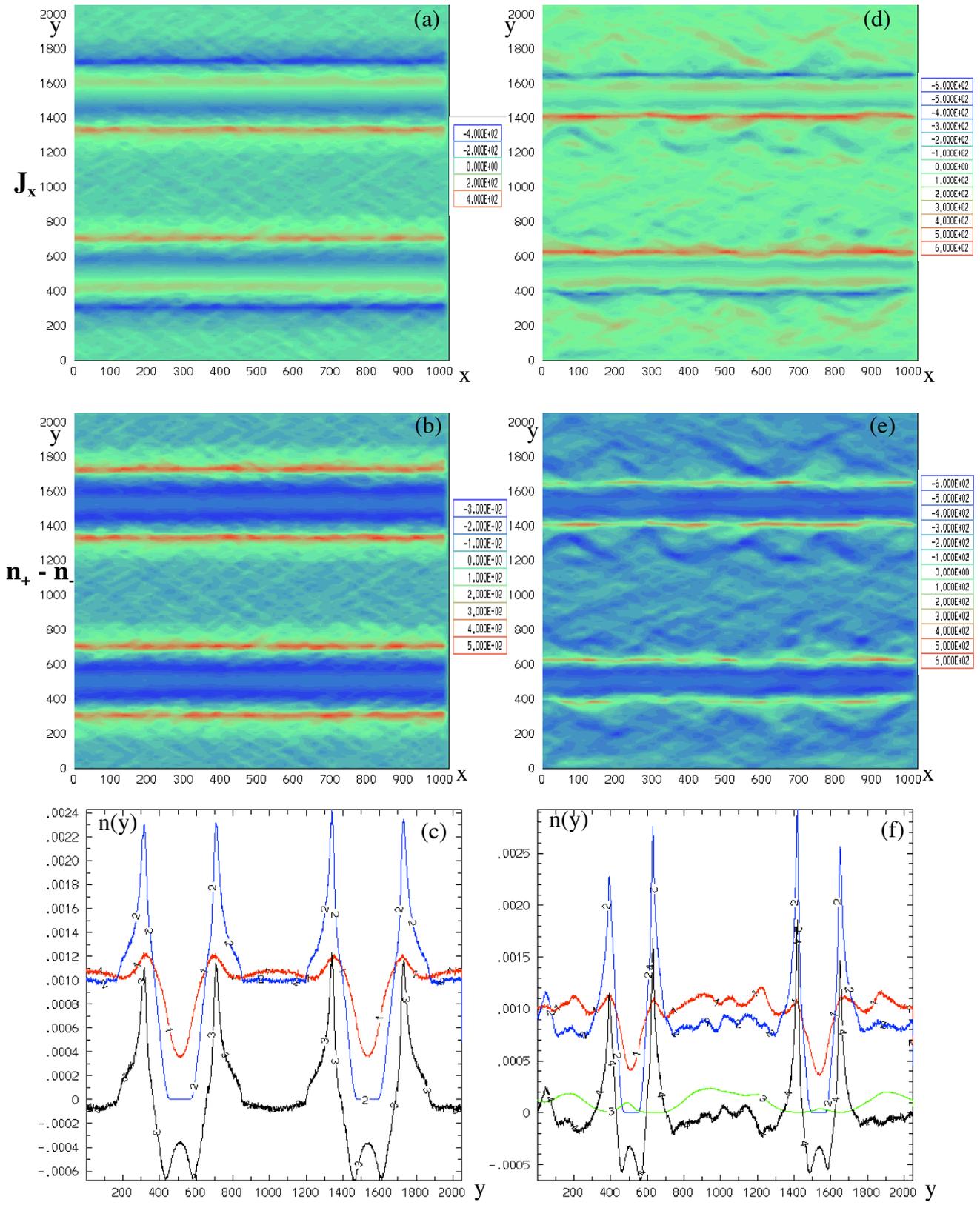

Fig.3

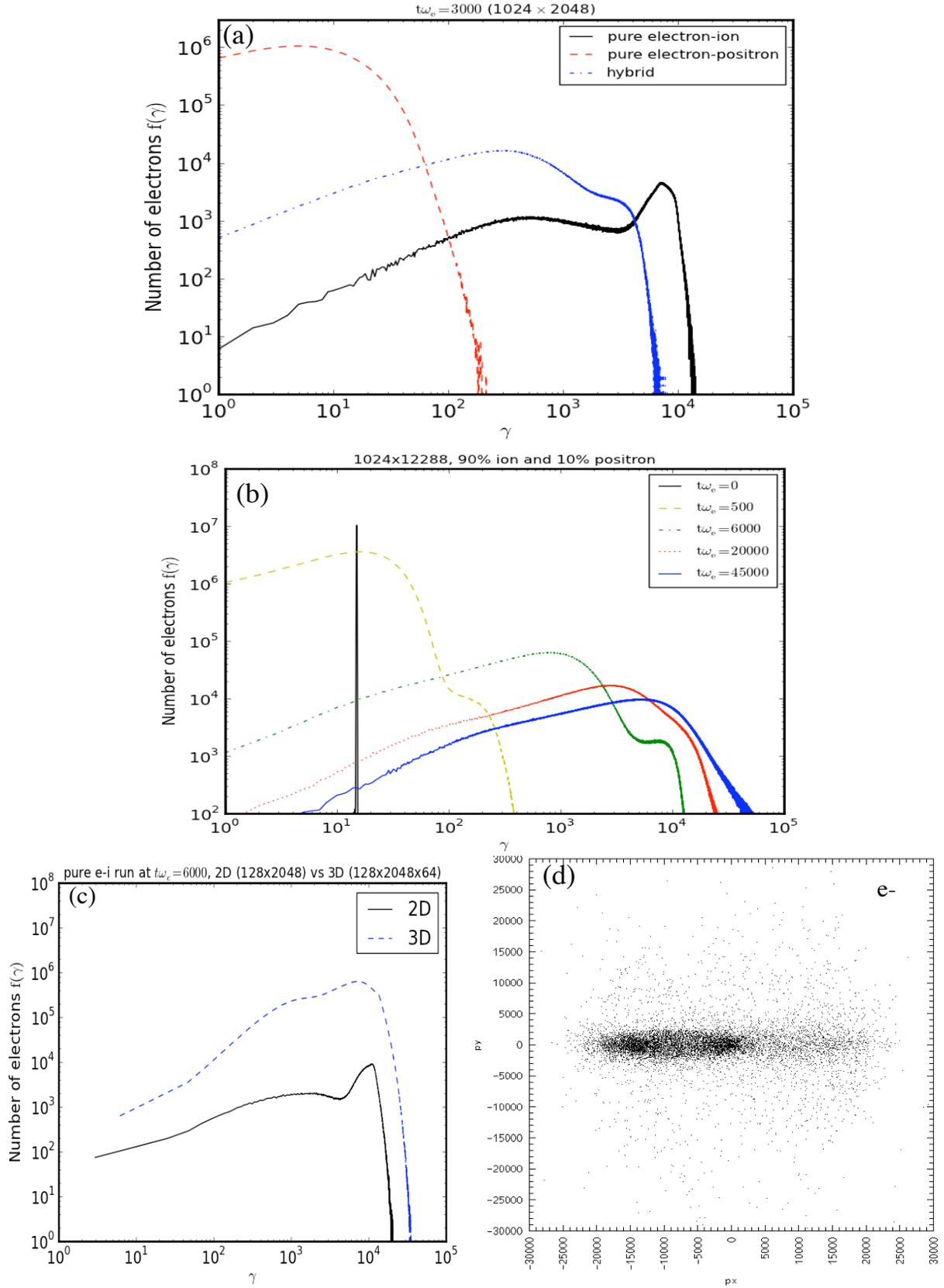

Fig.4

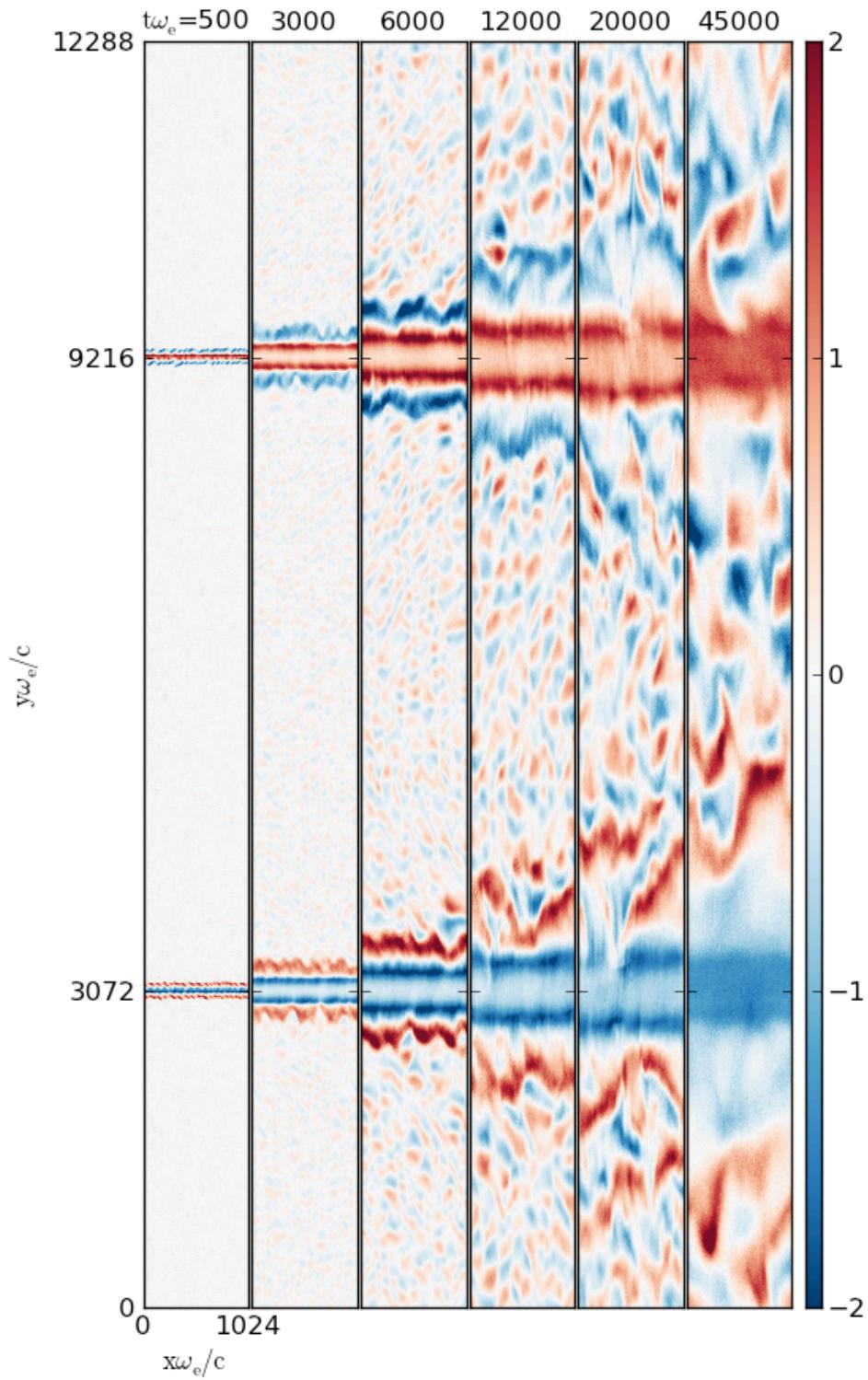

Fig.5
18